\documentclass{iopart}
\usepackage[dvips]{graphicx}

\def\MKMB{\textfont1=\textfont0\relax}\def\[#1]{$\MKMB #1$}
\def\lsim{\mathrel{\hbox{\lower .45ex\hbox{$\sim$}\llap{\raise .38ex\hbox{$<$}}}}}
\def\v#1{\mathchoice{{\mathord{\mbox{\boldmath$\displaystyle #1$}}}}{{\mathord{\mbox{\boldmath$\textstyle #1$}}}}{{\mathord{\mbox{\boldmath$\scriptstyle #1$}}}}{{\mathord{\mbox{\boldmath$\scriptscriptstyle #1$}}}}}
\def\gsim{\mathrel{\hbox{\lower .5ex\hbox{$\sim$}\llap{\raise.4ex\hbox{$>$}}}}}
\catcode`\@=11

\advance\textheight by 2.5cm
\advance\textwidth  by 1.0cm
\advance\voffset by -1.0cm
\advance\hoffset by -0.8cm

\begin{document}

\title{Novel electronic states close to
Mott transition in low-dimensional and frustrated systems}

\author{D. I. Khomskii}

\address{II~Physikalisches Institut, Universit\"at zu K\"oln,
Z\"ulpicher Str.~77, 50937 K\"oln, Germany}

\begin{abstract}
Recent studies demonstrated that there may appear
different novel states in correlated systems close to localized--itinerant
crossover.  Especially favourable conditions for that are met
in low-dimensional and in frustrated systems.
In this paper I discuss on concrete examples some of such novel states.
In particular, for some spinels and triangular systems there
appears a ``partial Mott transition'', in which first some finite clusters
(dimers, trimes, tetramers, heptamers) go over to the itinerant
regime, and the real bulk Mott transition occurs only later.
Also some other specific possibilities in this crossover regime are
shortly discussed, such as spin-Peierls--Peierls transition
in TiOCl, spontaneous charge disproportionation in some cases, etc.
\end{abstract}


\maketitle

\section{Introduction}
The subdivision of electronic states into delocalized, itinerant,
and localized, strongly correlated states is one of the most fundamental
ones in the description of solids.
Crudely, delocalized states exist when the intersite electron hopping
is larger than the typical electron--electron interaction
energy, and localized states exist in the opposite limit.
In quantum chemistry these two limits correspond e.g.\ to the description
of electrons in terms of molecular orbitals (Mulliken description,
known in chemistry as molecular orbital--linear combination of atomic orbitals
(MO--LCAO) method), and strongly correlated electrons are
described by the Heitler--London approximation.
For concentrated solids these two regimes give either
standard band systems (metals or semiconductors),
or Mott insulators.

The electron--electron repulsion, e.g.\ the Hubbard's $U$,
is essentially an atomic property
and does not depend on interatomic distance,
whereas the electron hopping strongly depends on this distance.
This leads to the concept, largely advocated e.g.\ by Goodenough \cite{goodenough},
that there should exist a critical interatomic distance~$R_c$
(different for different transition metal (TM) ions)
dividing the regions of localized and itinerant behaviour
of electrons.  When this distance changes, e.g.\ under pressure,
there should occur an insulator--metal (Mott) transition.

Usually Mott transition in TM compounds such as TM oxides
is supposed to occur homogeneously in the whole sample.
But, as became clear recently, this is not the only possibility.
There may appear novel, inhomogeneous states
close to a Mott transition, so that we have, in a sense,
a ``partial'' Mott transition: in some parts of the system,
in some particular clusters, the interatomic distances may
already be smaller than $R_c$, so that these clusters may be
better described using molecular orbitals,
whereas the distance {\it between} such clusters will still
remain large, so that there will be no net metallic
conductivity.  The formation of such ``metallic'' clusters
has much in common with the valence bond solids.

This phenomenon, i.e.\ such ``fractional'' Mott transition,
is most often observed in low-dimensional
and in frustrated systems. Indeed, in regular
three-dimensional lattices such as perovskites we usually
have homogeneous ordered states, e.g.\ 3d
antiferromagnetic ordering.  Molecular-type clusters
such as valence bond states are much more plausible in
low-dimensional and frustrated systems.
Below I will describe
this situation using several specific examples of systems
with triangular and pyrochlore (spinel) lattices,
and also on some quasi-one-dimensional systems.
I will also shortly mention some other possible
novel phenomena, which may appear in systems close to the
localized--itinerant crossover.

\section{Formation of dimers in spinels}
Metal--insulator transitions with the formation of unusual structures
are observed in many spinels.  The recent examples are
\[MgTi_2O_4] \cite{schmidt}, \[CuIr_2S_4] \cite{radaelli} or \[AlV_2O_4]~\cite{matsuno}.
Structural studies have shown that in all these cases rather unusual
structural modifications take place, which often can be described
as the formation of molecular clusters. In \[MgTi_2O_4] there appears
a ``chiral'' structure \cite{schmidt}, in \[CuIr_2S_4] Ir octamers are formed \cite{radaelli},
but both these phenomena can in fact be explained
by the formation of Ti or Ir {\it singlet dimers}, which in a
frustrated lattice of $B$-sites of a spinel finally give rise
to these chiral or octamer structures~\cite{mizokawa}.
And in \[AlV_2O_4] the molecular clusters formed consist of 7 vanadiums ---
real large ``heptamer molecules''.

\begin{figure}[ht]
  \centering
  \includegraphics[scale=1]{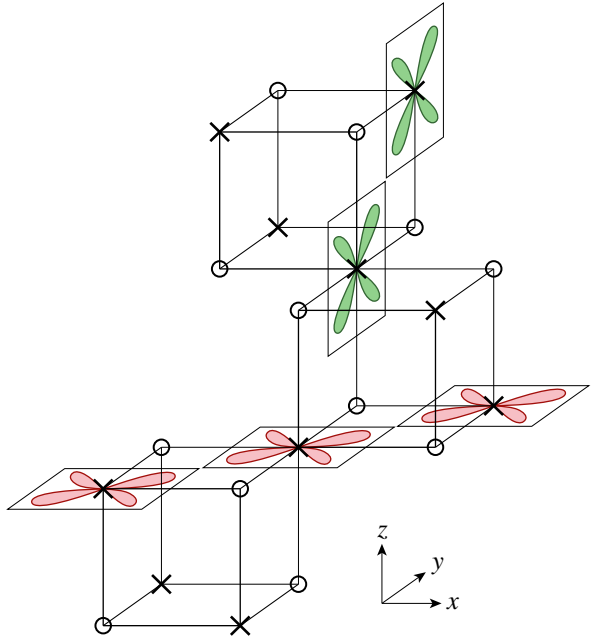}
  \caption{Formation of one-dimensional electronic spectrum for $t_{2g}$ electrons
     in a lattice of $B$-sites of a spinel.  The $xy$ and $yz$ orbitals are shown.}
  \label{fig:1}
\end{figure}

In all these cases the formation of these clusters is largely
caused by specific features of the $B$-site spinel
lattice, which may be represented as a collection of 1d chains
running in $xy$ (or $x\bar y$), $xz$ and $yz$ directions, see Fig.~\ref{fig:1}.
The $t_{2g}$-orbitals of Ti or V ions on this lattice
have a strong direct $d$--$d$ overlap, such that e.g.\
the electrons from the $xy$-orbitals can hop to the same orbitals
in neighbouring ions in the $xy$-direction, see Fig.~\ref{fig:1},
or similarly $yz$-electrons can hop in the $yz$-direction, etc.
As a result the electronic structure in these, basically cubic
crystals, has essentially one-dimensional character.  And
in the insulating phase one forms a Peierls state with
singlet dimers in these 1d chains~\cite{mizokawa}, which
for a spinel lattice give the chiral structure in \[MgTi_2O_4],
the heptamers in \[AlV_2O_4] and octamers (consisting actually of dimers)
in \[CuIr_2S_4].
Note that the interatomic distances in these clusters are rather short:
they
are shorter
than the critical distance $R_c$ for the localized--itinerant
crossover~\cite{goodenough}, so that one can consider
the electronic state in these clusters as forming molecular
orbitals, i.e.\ in these clusters we are already on the ``metallic'' side
of the Mott transition.  But the distance {\it between} these clusters
is larger than $R_c$ and as a result the total compound is insulating.

\begin{figure}[ht]
  \centering
  \includegraphics[scale=1]{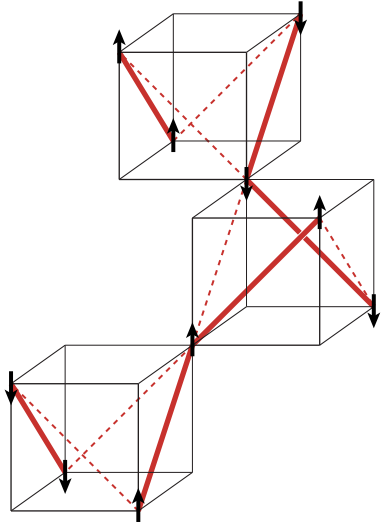}
  \caption{Formation of short (bold) and long (dashed) V--V bonds in \[ZnV_2O_4],
     following \cite{pardo}; the same pattern also exists in \[CdV_2O_4] \cite{giovanetti}.}
  \label{fig:2}
\end{figure}

The ``metallic'' clusters in the examples discussed above are singlet.
But this is not the only possibility. Magnetic (e.g.\ triplet)
clusters can also be formed in some cases.  This seems to happen
in \[ZnV_2O_4]~\cite{pardo}.  V spinels such as \[MgV_2O_4], \[ZnV_2O_4],
\[CdV_2O_4] show cubic--tetragonal structural transitions~\cite{furubayashi}
with the formation of a very specific magnetic structure at lower
temperatures, with ${\uparrow}\,{\uparrow}\,{\downarrow}\,{\downarrow}$
pattern in $xz$- and $yz$-chains (and with ${\uparrow}\,{\downarrow}\,{\uparrow}\,{\downarrow}$
ordering in the $xy$-direction)~\cite{reehuis}, see Fig.~\ref{fig:2}.
Usually the properties of these systems are explained by some type
of orbital ordering:
the $xy$-orbitals are always occupied due to tetragonal
distortion with $c/a<1$, and the remaining second $t_{2g}$ electron
of the \[V^{3+}]($d^2$) ions is supposed to form
an additional orbital ordering of some type~\cite{motome,tchernyshev,mizokawa}.
But our recent ab~initio calculations~\cite{pardo} have shown
that there should exist in \[ZnV_2O_4] a V dimerization
in $xz$- and $yz$-chains, see Fig.~\ref{fig:2}, with, surprisingly,
ferromagnetic bonds becoming shorter (thick bonds in Fig.~\ref{fig:2})!  One can explain this tendency
by noticing that, in contrast to most examples of valence
bond solids with singlet bonds, here we are dealing not with ions with
one electron with $S=\frac12$, but with $d^2$ ions with~$S=1$.
One can argue that the extra delocalization of one of two electrons
in a short bond enhances the ferromagnetic (actually double-exchange)
interaction between these $d^2$, $S=1$ ions.
The ab~initio calculations of another material of this family, \[CdV_2O_4],
done by a different method, have confirmed the formation of
these dimers, and demonstrated, in addition, that their
formation can lead to the appearance of ferroelectricity, which was
indeed observed in \[CdV_2O_4]~\cite{giovanetti}. 

Interestingly enough, short V--V distances obtained
theoretically for \[ZnV_2O_4] are only $2.92\,$\AA\ --- again
shorter than the critical V--V distance $2.94\,$\AA~\cite{goodenough}.
Thus again these V dimers may be considered as ``metallic'',
though the material itself remains insulating,
albeit with a rather small energy gap $\sim 0.2\,\rm eV$~\cite{pardo}.

\section{``Metallic'' clusters in layered materials}
\begin{figure}[ht]
  \centering
  \includegraphics[scale=1]{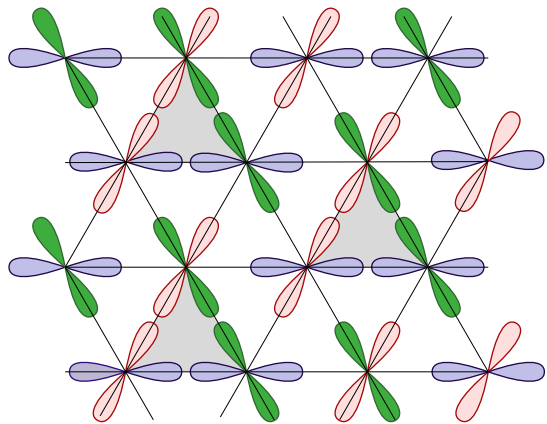}
  \caption{Orbital ordering in \[LiVO_2] with three orbital sublattices,
     leading to the formation of spin singlet states on tightly
     bound V triangles (shaded), following \cite{pen}.}
  \label{fig:3}
\end{figure}

Some more examples of ``metallic'' clusters in an overall
insulating matrix are found in layered materials, notably
with triangular lattices, such as \[LiVO_2] \cite{pen},
\[LiVS_2] \cite{katayama}, \[TiI_2]~\cite{meyer}.
This lattice is usually considered as frustrated,
meaning that it is not bipartite.
But in \[LiVO_2], \[LiVS_2], \[TiI_2] we are dealing with
ions \[Ti^{2+}], \[V^{3+}] with two $t_{2g}$ $d$-electrons, which,
besides spin, have also triple orbital degeneracy.
And, if one cannot subdivide a triangular lattice into two
sublattices, one can naturally subdivide it into three!
That is indeed what happens in these materials \cite{pen},
see Fig.~\ref{fig:3}, in which the occupied orbitals are shown.  After forming this orbital
superstructure, we have strong antiferromagnetic exchange
in the shaded triangles of Fig.~\ref{fig:3},
and in effect these bonds become shorter than the others, and there are singlet
states formed {\it on such trimers} (three $S=1$ ions with strong
antiferromagnetic coupling have a singlet ground state with $S_{\rm tot}=0$).
Our recent detailed ab~initio calculations \cite{hua wu} confirmed
this picture and gave a strong antiferromagnetic exchange
in a trimer, the exchange interaction between trimers being
much weaker and ferromagnetic.
Lattice optimization gave V--V distances in a trimer of $2.56\,$\AA,
in good agreement with the experimental value.
Note that in this case the short distance in a cluster --- here
a trimer --- is even shorter than the V--V distance in V metal, $2.62\,$\AA, i.e.\ these
clusters can indeed be considered as ``metallic''.

\begin{figure}[ht]
  \centering
  \includegraphics[scale=1]{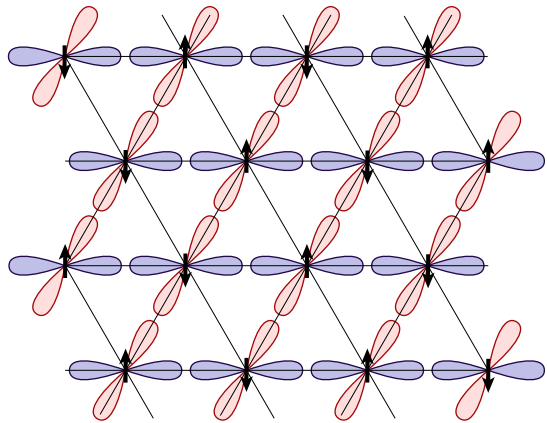}
  \caption{Orbital and magnetic ordering in \[NaVO_2] \cite{mcqueen}.}
  \label{fig:4}
\end{figure}

If however we increase the total lattice parameter,
which we can do e.g.\ by going from \[LiVO_2] to \[NaVO_2],
the situation changes drastically.
As shown in \cite{mcqueen}, the orbital ordering in \[NaVO_2]
is quite different from that in \[LiVO_2]: only two types
of orbitals are occupied, see Fig.~\ref{fig:4}.
As a result, after such orbital ordering the system becomes
topologically a square lattice, and as such it
can develop the ordinary N\'eel magnetic ordering,
which it actually does.
Note also that, in contrast to \[LiVO_2], in \[NaVO_2] the bonds
with strong orbital overlap 
become {\it not shorter, but longer\/}!  One can explain this
as a competition of two effects: stronger
covalency tends to make such bonds shorter,
but the bond--charge repulsion (Coulomb repulsion
of electron clouds on orbitals directed towards each other)
would tend to make such bonds longer.
Apparently when the system has a relatively small lattice
parameter, or small average TM--TM distance,
i.e.\ when the system is close to the localized--itinerant crossover,
the hybridization is rather strong, and the first effect,
the tendency to increase it still further, prevails,
so that such bonds become even shorter; this is what happens
in \[LiVO_2] and \[LiVS_2].
If, however, the system has localized $d$-electrons and
is far from the Mott transition, the $d$--$d$ overlap
is exponentially small, and the second effect,
the bond--charge repulsion, may become stronger,
as a result of which it is favourable to make such bonds
longer, to reduce this repulsion further.
This is apparently what happens in \[NaVO_2].

Until now we saw the formation of different types of TM clusters,
mostly singlet ones: dimers (\[MgTi_2O_4], \[CuIr_2S_4]),
trimers (\[LiVO_2], \[LiVS_2], \[TiI_2]),
or larger clusters such as heptamers in \[AlV_2O_4].
Also singlet tetramers can be formed, especially
in layered materials.
One such example is \[CaV_4O_9] \cite{starykh}, with the
interesting ``chiral'' lattice, in which the singlets
are formed by four V ions around an ``empty'' plaquette.
In all these cases either low-dimensionality or the frustrated nature
of the lattice apparently plays an important role.
As mentioned above, valence bond solids are indeed especially easily formed
in frustrated or quasi-one-dimensional systems --- this is in fact
the essence of the Peierls phenomena.

\section{Spin-Peierls-to-Peierls transition in TiOCl}
A very clear example of this phenomenon, in particular its features
close to the localized--itinerant crossover, is observed in TiOCl.
This quasi-one-dimensional material is insulating with localized $d$-electrons
at ambient pressure, and it has a spin-Peierls transition at $\sim 60\,\rm K$
(with an incommensurate phase at $60\,{\rm K} < T \lsim 90\,\rm K$).
Under pressure its energy gap strongly decreases,
and it approaches an insulator--metal transition at~$\sim 10\,\rm GPa$~\cite{kuntscher}.

\begin{figure}[ht]
  \centering
  \includegraphics[scale=1]{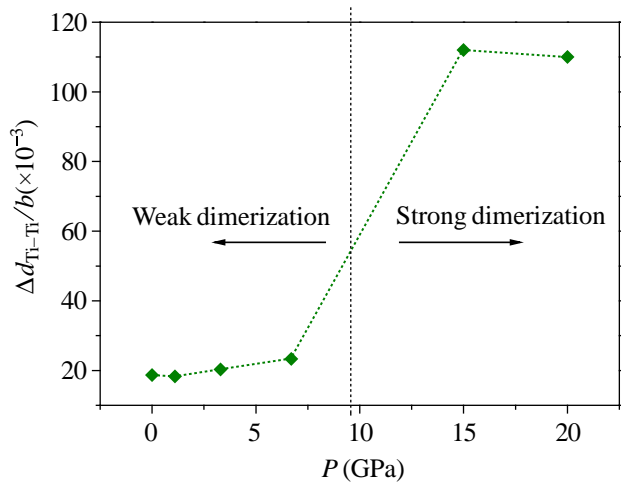}\qquad \null\par
  \caption{Change of dimerization in TiOCl with pressure \cite{forthaus}.}
  \label{fig:5}
\end{figure}

However direct dc resistivity measurements have shown \cite{forthaus} that TiOCl
actually remains insulating even for $P>10\,\rm GPa$,
at least up to $30\,\rm GPa$, albeit with a smaller gap.
Surprisingly, ab~initio calculations with lattice optimisation
have shown \cite{blanco-canosa} that in the high-pressure phase, in which
indeed the energy gap strongly decreases, the Ti--Ti dimerization
itself does not decrease but {\it increases} instead, see Fig.~\ref{fig:5}\thinspace!

A qualitative explanation can be obtained, using again
the idea of the proximity of this material to the localized--itinerant
crossover.
At $P=0$ the $d$-electrons are localized, and we have a spin-Peierls
transition, with $T_c$ and with the degree of dimerization $\delta d/d$
proportional to
\begin{equation}
T_c \sim \frac{\delta d}{d} \sim J\,e^{-1/\lambda}
\end{equation}
where $J$ is the exchange interaction, $J\sim t^2/U$, and $\lambda$
is the spin--phonon coupling constant.
Note that the energy gap in this regime is still of the Mott--Hubbard
type, i.e.\ it is large, $E_g\sim U$.
Under pressure TiOCl approaches
the insulator--metal transition, as observed in~\cite{kuntscher}.
But after crossing to the itinerant regime, the system still ``knows''
that it is one-dimensional, and as such it has a Peierls instability,
with $T_c$ ($\sim{}$degree of dimerization${}\sim{}$energy gap)
\begin{equation}
T_c \sim E_g \sim \frac{\delta d}{d} \sim t\,e^{-1/\lambda'}\;,
\end{equation}
with $\lambda'$ being the electron--electron coupling constant in this
regime.
That is, the very nature of insulating state
in this regime is {\it due to Peierls dimerization}, and is not
of Mott insulator type.  But the dimerization is in this regime
proportional to the bandwidth, or the hopping matrix element~$t$,
and not to $J\sim t^2/U$.  Thus apparently
we have here, simultaneously with the localized--itinerant crossover,
also a spin-Peierls--Peierls transition.

\section{Other possible specific phenomena close to a Mott transition}
As we have discussed above, a possible situation, especially in low-dimensional
and frustrated systems, is that the Mott transition occurs
not in the whole system simultaneously, but ``piecewise'', first
in some finite clusters such as dimers, trimers, tetramers etc.
But other nontrivial states may also occur in the vicinity of Mott transitions,
different both from those typical for strong Mott insulators and for normal
metals.  These situations can occur also for simple lattices
such as those of perovskites.  In this chapter for completeness
I will also shortly discuss some of these phenomena.

\subsection{Spontaneous charge disproportionation vs orbital ordering}
In some systems with strongly correlated electrons charge ordering occurs
with decreasing temperature.  Usually it appears in systems
with noninteger electron concentration (fractional number of
electrons per TM ion).
The well-known examples are the Verwey transition in magnetite
\[Fe_3O_4] or charge ordering in half-doped manganites
such as \[La_{0.5}Ca_{0.5}MnO_3],
see e.g.~\cite{imada}.
But there are also systems in which charge ordering, or rather
spontaneous charge disproportionation, appears in case of, formally,
integer electron occupation, or integer valence.  Such is
for example the situation in \[CaFeO_3], in which
there occurs charge segregation
\begin{equation}
2{\rm Fe}^{4+}(d^4) \Longrightarrow {\rm Fe}^{3+}(d^5) + {\rm Fe^{5+}}(d^3)\;.
\label{eq:3}
\end{equation}
A phase transition with the formation of a \[Fe^{3+}]--\[Fe^{5+}]
superstructure occurs in \[CaFeO_3] at $T_c=290\,\rm K$ and is simultaneously
a metal--insulator transition~\cite{takano}.
Note that a similar compound \[SrFeO_3] with somewhat broader
bands remains metallic down to lowest temperatures.
Apparently the phenomenon of charge disproportionation
in this case occurs in the system close to the localized--itinerant
crossover.

Similar spontaneous charge disproportionation occurs
also in \[La_{2/3}Sr_{1/3}FeO_3] \cite{takano} (\[3Fe^{4+} \Longrightarrow 2Fe^{3+} + Fe^{5+}]),
but also e.g.\ in nickelates $R$\[NiO_3], especially
those with small rare earths $R = \rm Lu$, Y, etc.~\cite{alonso,mizokawa2,mazin}.
In nickelates charge disproportionation may be
described by the ``reaction''
\begin{equation}
2{\rm Ni}^{3+}(t_{2g}^6e_g^1) \Longrightarrow {\rm Ni^{2+}}(t_{2g}^6e_g^2) + {\rm Ni}^{4+}(t_{2g}^6)
\end{equation}
(\[Ni^{3+}] and \[Ni^{4+}] here are in the low-spin state).
In these systems charge disproportionation is also a metal--insulator
transition.

Yet other systems showing
the same phenomenon
are materials formally containing \[Au^{2+}]($d^9$),
which at
normal conditions always disproportionate into
\[Au^{1+}]($d^{10}$)${}+{}$\[Au^{3+}]($d^8$).
This occurs e.g.\ in \[CsAuCl_3] which at ambient pressure
contains two inequivalent Au sites (its formula
is usually even written as \[Cs_2Au_2Cl_6]).
Under pressure this system also becomes metallic
with equivalent Au ions~\cite{matsushita}.

One can notice three specific features, which
apparently are important for the very phenomenon
of spontaneous charge disproportionation.
First, in TM compounds such charge disproportionation
occurs in systems in which the TM ion in the ``original''
state would have been a strong Jahn--Teller ion,
with double orbital degeneracy (\[Fe^{4+}]($t_{2g}^3e_g^1$);
low-spin \[Ni^{3+}]($t_{2g}^6e_g^1$); \[Au^{2+}]($t_{2g}^6e_g^3$)).
Thus it seems that this spontaneous charge disproportionation
appears in \[Fe^{4+}] and \[Ni^{3+}] compounds
{\it instead of} the Jahn--Teller or orbital ordering,
when the systems are close to localized--itinerant crossover~\cite{phys.scripta,mazin}.
But in  ``\[Au^{2+}]'' compounds such charge disproportionation
is stabilized by the {\it enhanced} JT distortion instead~\cite{ushakov}.
Second, the TM ionic states involved typically correspond to states
with unusually high oxidation states, or untypically high
valence (\[Fe^{4+}] and \[Fe^{5+}]; \[Ni^{3+}] and \[Ni^{4+}]; \[Au^{3+}]).
As follows from the standard classification~\cite{bocquet},
these states correspond to the situation with small or negative
charge-transfer gaps and with a large contribution of
ligand (e.g.\ oxygen) $p$-holes.  In effect e.g.\ the reaction~(\ref{eq:3})
should be rather written as
\begin{equation}
2{\rm Fe}^{3+}\underline L \Longrightarrow {\rm Fe}^{3+} + {\rm Fe}^{3+}\underline L^2\;,
\label{eq:5}
\end{equation}
where $\underline L$ denotes the ligand hole
(e.g.\ \[Fe^{3+}]$\underline L$ instead of \[Fe^{4+}]
denotes the state \[Fe^{3+}]($d^5$)\[O^-]($2p^5$)).
Apparently the large contribution of ligand holes facilitates the
process of charge disproportionation (the increase of Coulomb
repulsion in the reaction~(\ref{eq:5}) is definitely much
smaller than if it would occur on the $d$-shells~(\ref{eq:3})).

These factors lead to the third conclusion: that the
spontaneous charge disproportionation is typically
observed in systems close to an insulator--metal transition.
Experimentally this is indeed the case: in the examples
mentioned above such disproportionation coincides with
metal--insulator transition.

\subsection{Phase separation}
Without going into details, let us shortly mention yet
another specific phenomenon, which is typically observed
close to metal--insulator transitions.
This is the phenomenon of phase separation, which often occurs
especially in doped Mott insulators.
It is found experimentally in many cases;
in manganites~\cite{moreo,khomskii,babushkina}; in cobaltites~\cite{kuhns};
and apparently also in cuprates~\cite{gorkov}.
The tendency towards phase separation is also seen
theoretically in many models: in the Hubbard \cite{visscher}
and in the $t$--$J$ models~\cite{emery}; in the double-exchange
models \cite{kagan}, etc.
The resulting state may be visualised as a random
system with metallic regions in an insulating matrix.  The properties
of such state can be described by
percolation theory --- a classical description
of random systems, although quantum effects in this case can also
be very important.

\subsection{Large spontaneous currents in frustrated systems}
Typical for many frustrated systems, which
contain triangles or tetrahedra as building blocks,
is an extra degeneracy, which can be characterized
by spin chirality.  Thus the ground state
of an equilateral triangle of $S=\frac12$ spins
with an antiferromagnetic interaction is a quartet,
with the total spin $S_{\rm tot}=\frac12$ and with extra
double degeneracy, which can be characterized by the scalar
spin chirality
\begin{equation}
\chi_{123} \sim \v S_1 \cdot [\v S_2 \times \v S_3]\;.
\label{eq:6}
\end{equation}
The physical meaning of the scalar chirality was
clarified in \cite{bulaevskii} (see also \cite{khomskii2}), where it was shown
that the states of a triangle with nonzero chirality~(\ref{eq:6})
carry a spontaneous circular electric current $j\sim\chi_{123}$.
Correspondingly, such triangle would have an orbital magnetic
moment $M\sim j\sim \chi_{123}$.  In the nondegenerate
Hubbard model
with strong interaction $U\gg t$ the expression for this current,
obtained in perturbation theory in $t/U\ll 1$, has the form
\begin{equation}
j = \frac{24et^3}{\hbar U^2}(\v S_1\cdot[\v S_2\times\v S_3])\;.
\end{equation}
In this limit the current and the corresponding orbital moment are small.
But one can show on general grounds, by using symmetry arguments,
that such currents would be proportional to the chirality~(\ref{eq:6})
in the general case, for $t\gsim U$, and also close to Mott transition.
In such situations the corresponding current and orbital
moments can be quite large.  Indeed, estimates done in \cite{georgeot},
using the exact solution for a triangle,
show that for reasonable values of parameters
the orbital magnetic moments can be as large as $\sim 0.7\mu_B$ ---
comparable with the spin moment.  The authors of \cite{georgeot}
suggested to use this degree of freedom for quantum computations.
This is yet another nontrivial property of frustrated systems,
which can lead to large and potentially useful effects close to Mott
transitions.

\section{Conclusion}
In conclusion, I want to stress once again that the systems
close to Mott transition, or to a localized--itinerant crossover,
can show a variety of specific properties, different from
those of states deep in the Mott insulator regime or those
of conventional metals.  Especially interesting is the situation
of a ``partial Mott transition'', in which first certain finite
clusters go over to the regime with uncorrelated (``itinerant'')
electrons, and only later there occurs a real Mott transition
in the whole sample.  As I tried to show on many examples,
the most favourable conditions for this phenomenon are met
in low-dimensional and in frustrated systems;
they are less probable in simple lattices such as perovskites.
But also other specific phenomena, less sensitive to the type of the
lattice, such as spontaneous charge disproportionation, can
occur close to insulator--metal transitions.
Thus this crossover region between localized and itinerant
electronic states presents a rich playground for interesting physics,
with eventually some novel states appearing in this situation.

\medskip

I am grateful to many people who contributed to works on which
this paper is based, especially M.~Abd-Elmeguid, C.~Batista,
L.~Bulaevskii, Hua Wu, I.~Mazin, M.~Mostovoy, V.~Pardo, R.~Rivadulla,
S.~Streltsov and A.~Ushakov.
This work was supported by the German program SFB~608 and by
the European project SOPRANO\null.


\section*{References}

\begin{thebibliography}{99}
\bibitem{goodenough}J.~B.~Goodenough, {\em Magnetism and the Chemical
Bond}, Wiley Interscience, New York, 1963

\bibitem{schmidt}M.~Schmidt et al., Phys. Rev. Lett. {\bf 92}, 056402 (2004)

\bibitem{radaelli}P.~G.~Radaelli et al., Nature {\bf 416}, 155 (2002)

\bibitem{matsuno}K.~Matsuno et al.,
J. Phys. Soc. Japan {\bf 70}, 1456 (2001)

\bibitem{mizokawa}D.~I.~Khomskii and T.~Mizokawa, Phys. Rev. Lett. {\bf 94}, 156402 (2005)

\bibitem{pardo}V.~Pardo et al., Phys. Rev. Lett. {\bf 101}, 256403 (2008)

\bibitem{furubayashi}T.~Furubayashi et al., J. Phys. Soc. Japan {\bf 66}, 778 (1994)

\bibitem{reehuis}M.~Reehuis et al., Eur. Phys. J. {\bf B35}, 311 (2003)

\bibitem{motome}Y.~Motome and H.~Tsunetsugu, Phys. Rev. B {\bf 70}, 184427 (2004)

\bibitem{tchernyshev}O.~Tchernyshev, Phys. Rev. Lett. {\bf 93}, 157206 (2004)

\bibitem{giovanetti}G.~Giovannetti et al., arXiv:1009.4077
(2010); Phys. Rev. B -- Rapid Comm. in press

\bibitem{pen}H.~Pen et al., Phys. Rev. Lett.
{\bf 78}, 1323 (1997)

\bibitem{katayama}N.~Katayama et al.,
Phys. Rev. Lett. {\bf 103}, 146405 (2009)

\bibitem{meyer}G.~Meyer, T.~Gloger and J.~Beekhuizen,
Z. Anorg. Allg. Chemie {\bf 635}, 1497 (2009)

\bibitem{hua wu}Hua Wu and D.~I.~Khomskii, to be publ.

\bibitem{mcqueen}T.~M.~McQueen et al., Phys. Rev. Lett. {\bf 101}, 166402 (2008)

\bibitem{starykh}O.~A.~Starykh et al.,  Phys. Rev. Lett. {\bf 77},
2558 (1996)

\bibitem{kuntscher}C.~A.~Kuntscher et al., Phys. Rev. B {\bf 74}, 184402 (2006);
Phys. Rev. B {\bf 76}, 241101 (2007)

\bibitem{forthaus}M.~K.~Forthaus et al., Phys. Rev. B {\bf 77}, 165121 (2008)

\bibitem{blanco-canosa}S.~Blanco-Canosa et al.,  Phys. Rev.
Lett. {\bf 102}, 056406 (2009)

\bibitem{imada}M.~Imada, A.~Fujimori and Y.~Tokura, Rev. Mod. Phys. {\bf 70} (1998), 1039

\bibitem{takano}M.~Takano et al., J. Solid State Chem. {\bf 39}, 75 (1981)

\bibitem{alonso}J.~A.~Alonso et al., Phys. Rev. Lett. {\bf 82}, 3871 (1999)

\bibitem{mizokawa2}T.~Mizokawa, D.~I.~Khomskii and G.~A.~Sawatzky,  Phys. Rev. B {\bf 61},
11263 (2000)

\bibitem{mazin}I.~I.~Mazin et al., Phys. Rev. Lett. {\bf98}, 176406 (2007)

\bibitem{matsushita}N.~Matsushita et al., J. Solid State Chem. {\bf 180}, 1353 (2007)

\bibitem{phys.scripta}D.~I.~Khomskii,
Physica Scripta {\bf 72}, CC8--CC14 (2005)

\bibitem{ushakov}A.~V.~Ushakov, S.~V.~Streltsov and D.~I.~Khomkii, to be publ.

\bibitem{bocquet}A.~E.~Bocquet et al., Phys. Rev. B {\bf 45}, 1561 (1992);
Phys. Rev. B {\bf 46}, 3771 (1992)

\bibitem{moreo}A.~Moreo, S.~Yunoki, and E.~Dagotto, Science {\bf 283},
2034 (1999)

\bibitem{khomskii}D.~I.~Khomskii, Physica B {\bf 280}, 325 (2000)

\bibitem{babushkina}N.~A.~Babushkina et al., J.
Appl. Phys. {\bf 83}, 7389 (1998)

\bibitem{kuhns}P.~L.~Kuhns et al,, Phys. Rev.
Lett. {\bf 91}, 127202 (2003)

\bibitem{gorkov}L.~P.~Gor'kov and A.~Sokol, JETP Lett. {\bf 46}, 420 (1987)

\bibitem{visscher}P.~B.~Visscher, Phys. Rev. B {\bf 10}, 943 (1974)

\bibitem{emery}V.~J.~Emery, S.~A.~Kivelson, and H.~Q.~Lin, Phys. Rev. Lett. {\bf 64}, 475 (1990)

\bibitem{kagan}M.~Yu.~Kagan, D.~I.~Khomskii, and A.~V.~Mostovoy, Eur.
Phys. J. {\bf 12}, 217 (1999)

\bibitem{bulaevskii}L.~N.~Bulaevskii, C.~D.~Batista, M.~V.~Mostovoy and D.~I.~Khomskii,
Phys. Rev. B {\bf 78}, 028402 (2008)

\bibitem{khomskii2}D.~I.~Khomskii, J. Phys. Cond. Matter
{\bf 22}, 164209 (2010)

\bibitem{georgeot}B.~Georgeot and F.~Mila, Phys. Rev. Lett {\bf 104}, 100502 (2010)








\end{thebibliography}
\end{document}